\documentclass[journal]{IEEEtran}
\usepackage{booktabs} 
\usepackage{amsmath}
\usepackage{graphicx}
\usepackage{subfigure}
\usepackage{makecell}
\usepackage[colorlinks, linkcolor=red, citecolor=blue]{hyperref}
\usepackage{amsfonts}
\usepackage{subcaption}
\usepackage{fontawesome} 

\ifCLASSINFOpdf
\else
\fi
\begin{document}
%
\title{Discriminating retinal microvascular and neuronal differences related to migraines: Deep Learning based Crossectional Study}

%
%
%

\author{Feilong Tang, Matt Trinh, Annita Duong, Angelica Ly, Fiona Stapleton FTSE, Zhe Chen,~\IEEEmembership{Member,~IEEE,} \newline Zongyuan Ge\textsuperscript{\faEnvelope}, Imran Razzak\textsuperscript{\faEnvelope} ~\IEEEmembership{Senior,~IEEE,}

 \thanks{F. Tang, Z. Chen and Z. Ge are with the  Faculty of Engineering, Monash University, Australia}
 \thanks{M. Trinh, A. Duong, A. Ly, and F. Stapleton are with School of Optometry \&	Vision Science and Centre for Eye Health, UNSW Sydney, Australia}
 \thanks{I. Razzak is with School of Computer Science and Engineering, University of New South Wales, Sydney, Australia}}

%
%

\markboth{Journal of \LaTeX\ Class Files,~Vol.~14, No.~8, August~2015}%
{Shell \MakeLowercase{\textit{et al.}}: Bare Demo of IEEEtran.cls for IEEE Journals}
%



\maketitle

\begin{abstract}
Migraine, a prevalent neurological disorder, has been associated with various ocular manifestations suggestive of neuronal and microvascular deficits. However, there is limited understanding of the extent to which retinal imaging may discriminate between individuals with migraines versus without migraines. In this study, we apply convolutional neural networks to color fundus photography (CFP) and optical coherence tomography (OCT) data to investigate differences in the retina that may not be apparent through traditional human-based interpretations of retinal imaging. Retrospective data of CFP type 1 [posterior pole] and type 2 [optic nerve head (ONH)] from 369 and 336 participants respectively were analyzed. All participants had bilaterally normal optic nerves and maculae, with no retinal-involving diseases. CFP images were concatenated with OCT default ONH measurements, then inputted through three convolutional neural networks – VGG-16, ResNet-50, and Inceptionv3. The primary outcome was performance of discriminating between with migraines versus without migraines, using retinal microvascular and neuronal imaging characteristics. Using CFP type 1 data, discrimination (AUC [95\% CI]) was high (0.84 [0.8, 0.88] to 0.87 [0.84, 0.91]) and not significantly different between VGG-16, ResNet-50, and Inceptionv3. Using CFP type 2 [ONH] data, discrimination was reduced and ranged from poor to fair (0.69 [0.62, 0.77] to 0.74 [0.67, 0.81]). OCT default ONH measurements overall did not significantly contribute to model performance. Class activation maps (CAMs) highlighted that the paravascular arcades were regions of interest. The findings suggest that individuals with migraines demonstrate microvascular differences more so than neuronal differences in comparison to individuals without migraines.

\end{abstract}


\IEEEpeerreviewmaketitle

\section{Introduction}

\IEEEPARstart{M}{igraines} can significantly disrupt quality of life, ranging from mild psychosomatic symptoms to severely debilating visual disturbances. It is the third most common disease and one of the most prevalent neurological disorders worldwide, affecting over 15\% of the global population~\cite{abdellatif2018effect}.  The Global Burden of Disease study ranks migraine as the leading cause of disability at the fourth level of the GBD scale~\cite{blair2023rimegepant}. Traditional migraine diagnosis involves assessing several factors, including the frequency and duration of headaches, the type of pain, the presence of visual auras, and family history of migraines~\cite{eigenbrodt2021diagnosis}. These are inherently subjective descriptions, which can vary widely~\cite{eigenbrodt2021diagnosis}. While studies indicate the potential of electroencephalography (EEG) in detecting migraines and identifying characteristic EEG markers, it requires substantial time for recordings (20 minutes for awake sessions and 30 minutes for sleep sessions)~\cite{craciun2014long} and suffers from low specificity~\cite{goker2023automatic}. 

\begin{figure}%
\centering
\subfigure[][]{%
\label{fig:ex3-a}%
\includegraphics[height=1.4in]{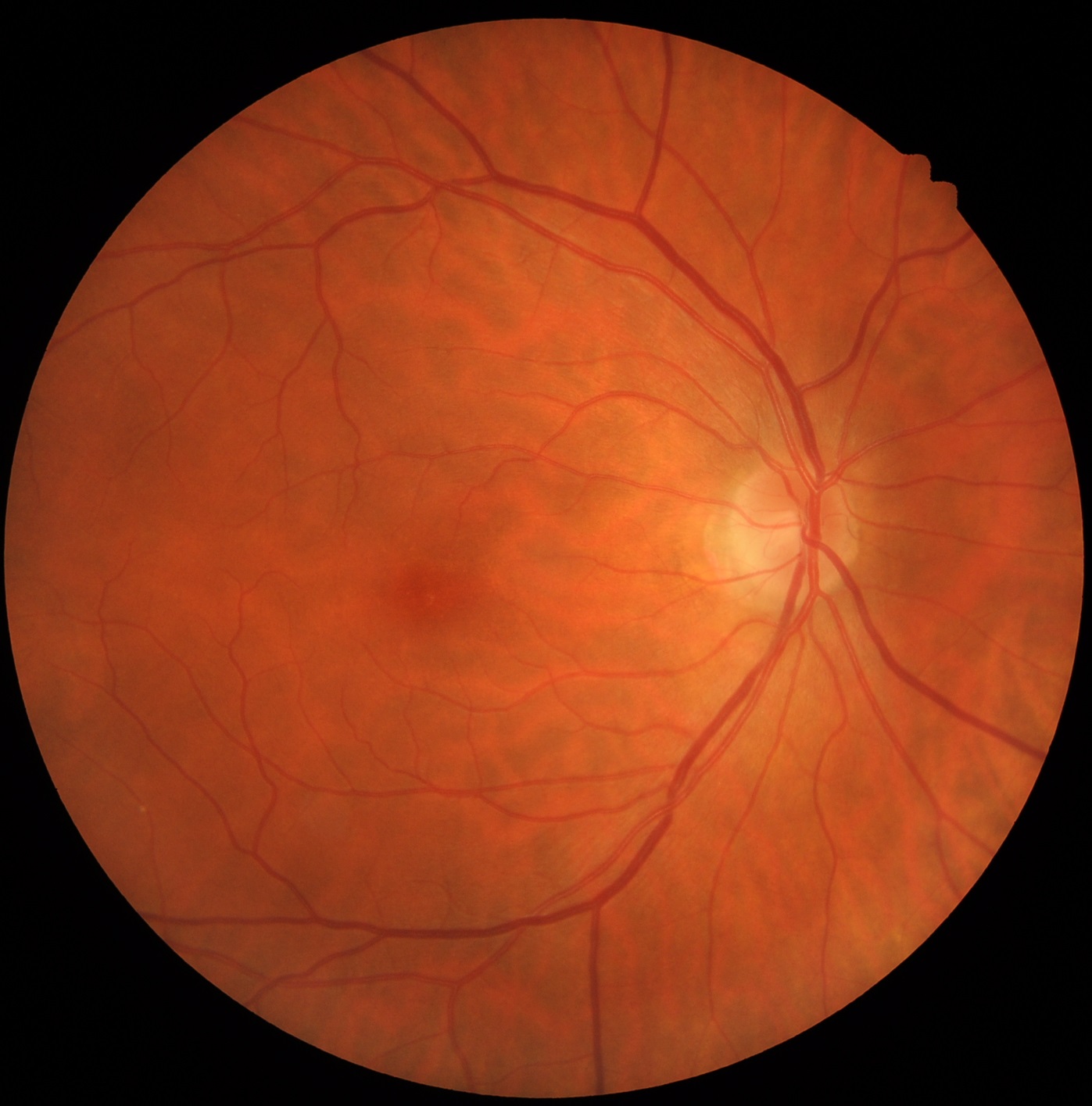}}%
\hspace{8pt}%
\subfigure[][]{%
\label{fig:ex3-b}%
\includegraphics[height=1.4in]{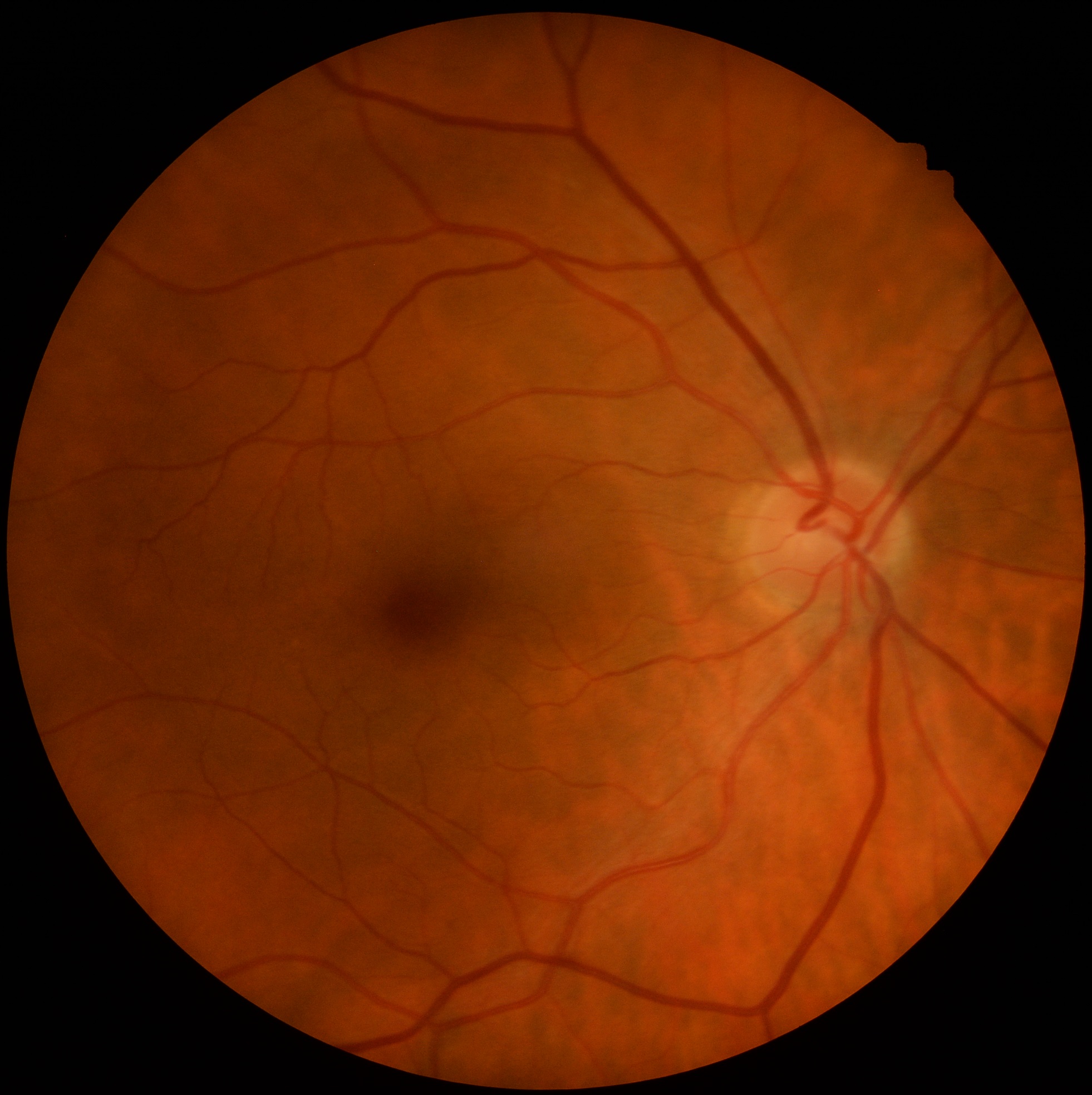}} \\
\subfigure[][]{%
\label{fig:ex3-c}%
\includegraphics[height=1.78in]{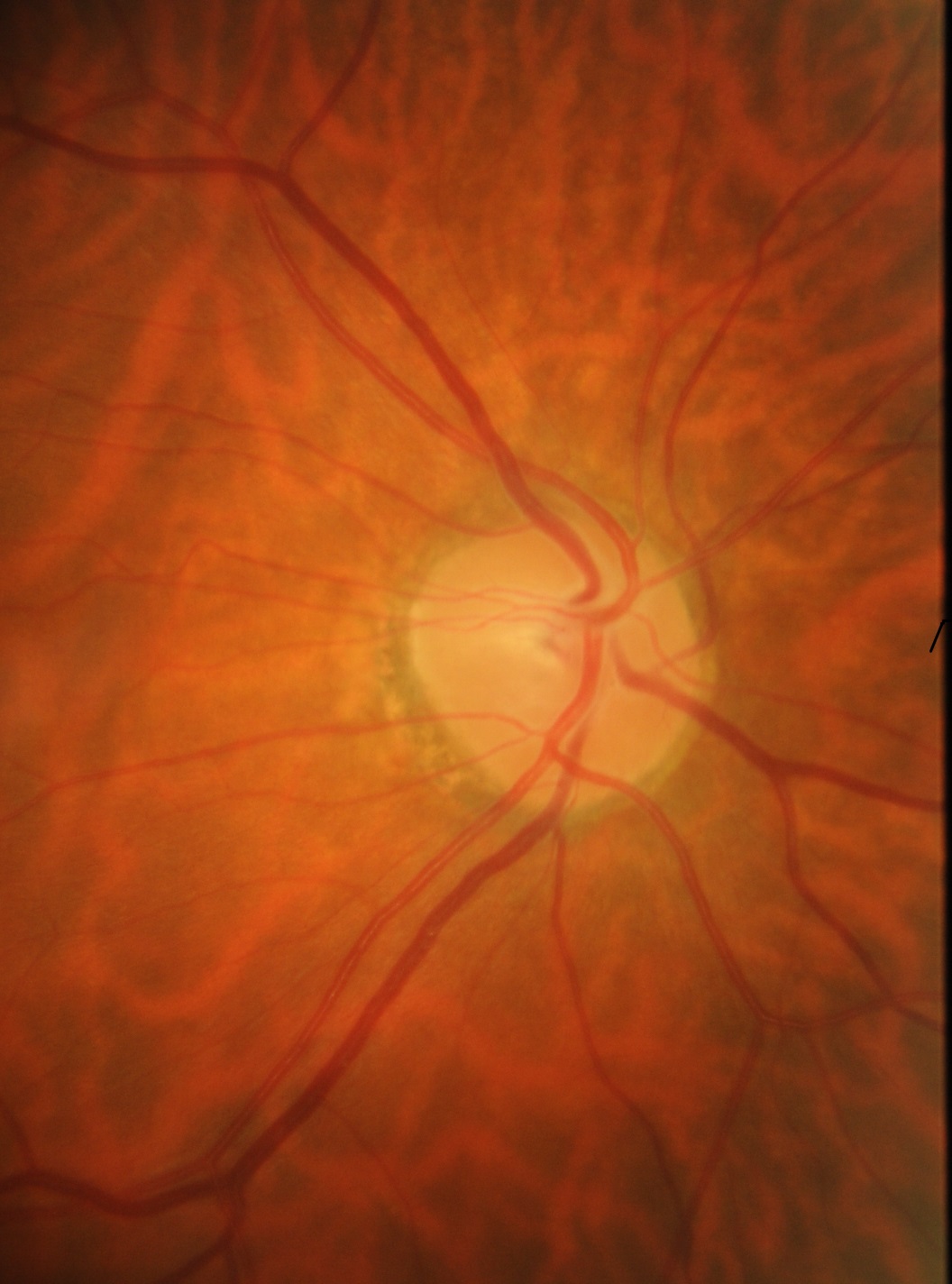}}%
\hspace{8pt}%
\subfigure[][]{%
\label{fig:ex3-d}%
\includegraphics[height=1.78in]{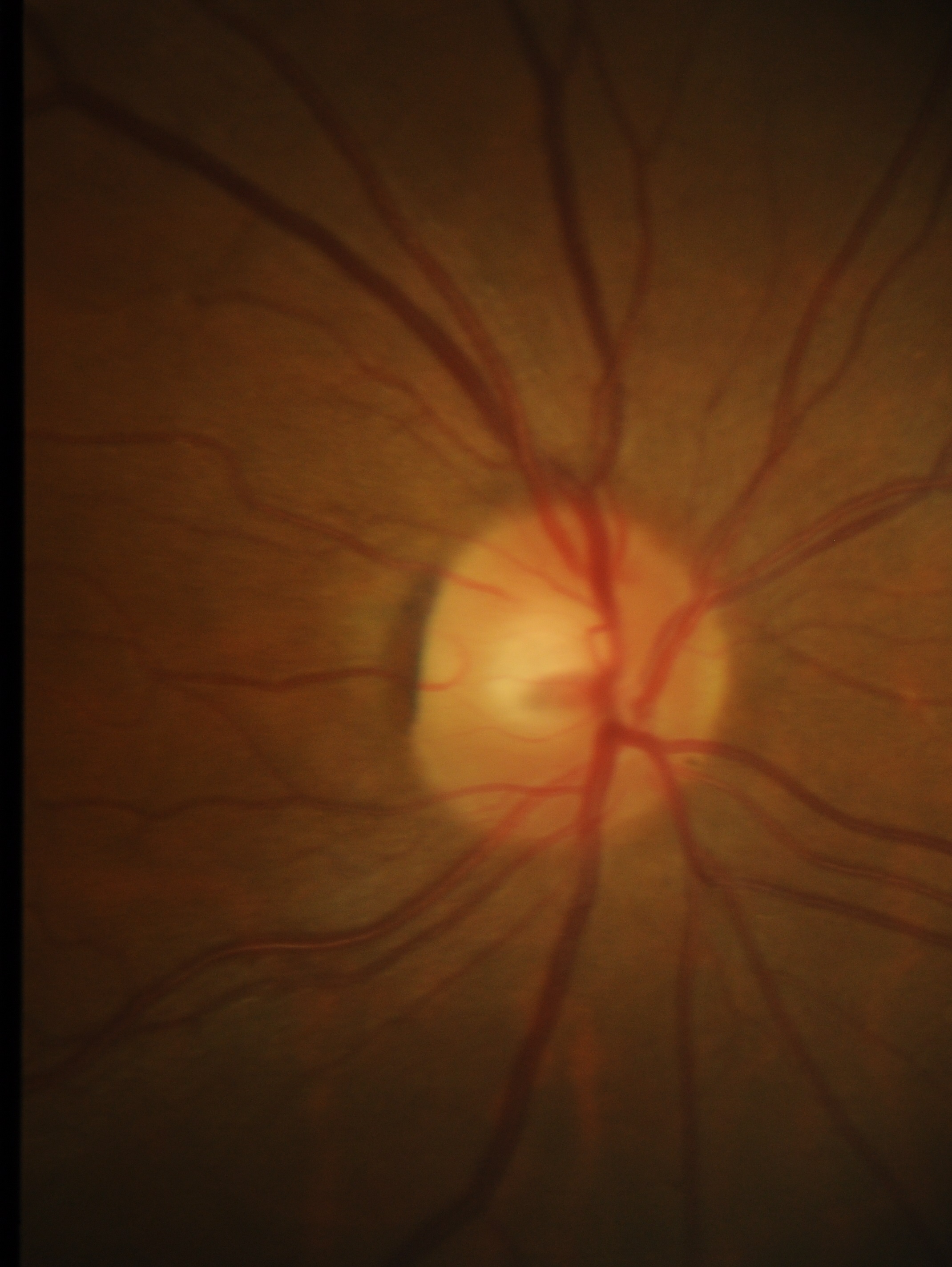}}%
\caption[.]{ Fundus photo of right eye (migraine vs healthy); Posterior Pole : 
\subref{fig:ex3-a}   Migraine;
\subref{fig:ex3-b} Healthy;
Optic Nerve Head
\subref{fig:ex3-c} Migraine,
\subref{fig:ex3-d} Healthy}%
\label{fig:example}%

\end{figure}


Migraine pathophysiology is complex and involves a mix of biological and psychological factors~\cite{ashina2019migraine}. The vascular hypothesis of migraine pathophysiology~\cite{ray1940experimental} posits that vasodilation is a key factor in migraine-related symptomatology. Increased levels of calcitonin gene-related peptide (CGRP) can trigger vasodilation, causing a surge in blood flow, which is thought to contribute to the throbbing pain of migraine headaches~\cite{ailani2021atogepant, haanes2023atogepant, lin2021retinal, koruga2024impact}. Additionally, recent research has linked migraine pathophysiology to the activation and sensitization of the neuronal system. Figure \ref{fig:example} visualizes the fundus photos (posterior pole and optic nerve head) of right eye of participant with and and without migraine. Several studies suggest that retinal nerve fibre layer (RNFL) thicknesses in all quadrants (superior, inferior, nasal, and temporal) are significantly thinned in chronic migraine patients~\cite{abdellatif2018effect,labib2020retinal,feng2016retinal,reggio2017migraine}, but otherwise differences may be subtle. Considering that the retina is an extension of the central nervous system, using non-invasive retinal imaging, such as color fundus photography (CFP) and optical coherence tomography (OCT), could provide valuable insights into disease pathophysiology \cite{khan2023retinal,akinniyi2023multi}. CFP enables superficial visualization of the posterior pole, including both neuronal and microvascular elements, while OCT facilitates depth-resolved and quantifiable resolution of these retinal elements~\cite{feng2016retinal,marziani2009evaluation}. By analyzing these measures, the impact of migraines on the retinal health can be better understood.



The application of machine learning, a subset of artificial intelligence, could be integrated alongside the use of retinal imaging for migraine assessment. Given the complexity of migraine involving both microvascular and neurological changes, there are significant challenges in creating a generalized model using traditional machine learning methods that utilize handcrafted features \cite{gel2024changes,trinh2024sight}. To address these limitations, advanced techniques such as convolutional neural networks (CNNs) - including ResNet~\cite{he2016deep}, VGG-16~\cite{simonyan2014very}, and Inceptionv3~\cite{szegedy2016rethinking} - can be applied to learn complex patterns from large retinal imaging datasets without explicit feature engineering.

In this study, we utilized CFP and OCT data to train and test three models: VGG-16, ResNet-50, and Inceptionv3. The primary aim was to distinguish between individuals with and without migraines. Additionally, this discrimination task provides valuable clues for future pathophysiological research, highlighting areas of interest that could lead to the development of improved monitoring tools.

\begin{table}[!htb]
\centering
\caption{Metrics for ONH Measurement}
\begin{tabular}{p{4cm} p{3.5cm}}
\hline
\textbf{Metric}  \\
\hline
Quadrant-1 (T) & Quadrant-2 (S) \\
Quadrant-3 (N) & Quadrant-4 (I) \\
Clock Hour 1-12 & RNFL Symmetry \\
Disc Area & Rim Area \\
Average C/D Ratio & Vertical C/D Ratio \\
Average RNFL Thickness& Cup Volume \\
\hline
\end{tabular}
\label{onh_metrics}
\end{table}

\section{Method}
\subsection{Dataset collection}

\begin{figure*}[!htb]
    \centering
    \includegraphics[width=\linewidth]{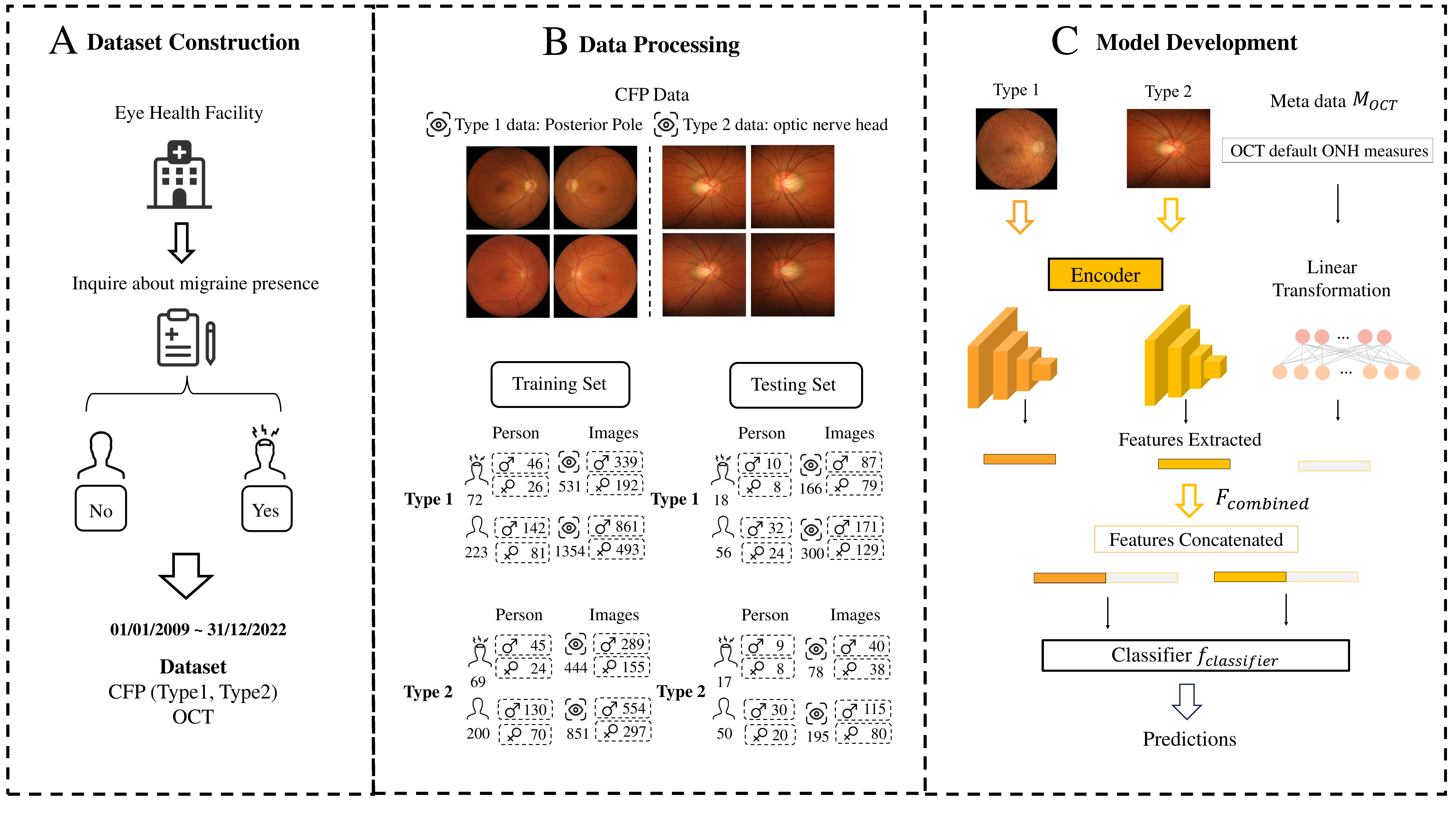}
    \caption{(A) Dataset Construction: The study used data from the Centre for Eye Health. The data were derived from a practical scenario in routine clinical practice.
    (B) Data Processing: CFP were divided into Type 1 data (posterior pole) and Type 2 data (optic nerve head). Data was further split into training and testing sets.
    (C) Model Development: Features were extracted from both CFP data types and OCT measurements to discriminate between migraine and non-migraine subjects.}
\label{fig1}
\end{figure*}

A consecutive sample of participants seen between 01/01/2009 to 31/12/2022 at the Centre for Eye Health, confirmed by two unblinded clinicians and author MT to have bilaterally normal optic nerves and maculae without any retinal disease. Data was collected from retrospective clinical records and patient self report (see Figure \ref{fig1}.(A)). All participants provided written informed consent for research use of their de-identified data, approved by the Biomedical Human Research Ethics Advisory Panel of the University of New South Wales and conforming to the tenets of the Declaration of Helsinki. Each participant may have multiple records included in the data. Temporal data from 3180 CFP images and 2679 OCT measurement data from both eyes of 369 participants acquired over 3033 visits (mean 8.22 visits per participant) were utilised to train the models. Temporal data from 739 CFP images and 598 OCT measurement data from both eyes of 72 migraine sufferers and 223 healthy controls, with no history of ocular or neurological disease, were utilised to test the diagnostic performance of the models. 


Two distinct types of CFP data were utilised to investigate retinal changes associated with migraines, as demonstrated in Fig.~\ref{fig1}(B). CFP Type 1 data comprised of low magnification images of the posterior pole. These images capture a 45° view of the macula, vascular arcades, and optic nerve head, which are often analyzed for microvascular changes and abnormalities. CFP Type 2 data consisted of high magnification images focusing specifically on the optic nerve head (ONH) and peripapillary retinal nerve fiber layer (RNFL).

Beyond CFP data, each category contained relevant clinical textual data (OCT quantitative), with 23 attributes per image, allowing for more in-depth analysis. OCT can precisely measure optic nerve head and peripapillary cross-sectional images of the retinal layers. The specific optic nerve head and RNFL metrics are shown in Table~\ref{onh_metrics}.
The quadrant measurements (Temporal, Superior, Nasal, Inferior) and Clock Hour 1-12 segments provide detailed thickness data for the retinal nerve fiber layer (RNFL) around the optic nerve head. 

\begin{table*}[!htb]
\centering

\caption{Distribution of CFP type 1 data [posterior pole]}
\begin{tabular}{lcccc}
\hline
 & \multicolumn{2}{c}{Train} & \multicolumn{2}{c}{Test} \\
\cline{2-5}
  & Migraine & No Migraine & Migraine & No Migriane \\
\hline
Number of Persons & 72 & 223 & 18 & 56 \\
Number of Images & 531 & 1354 & 166 & 300 \\
Age & \makecell{55.2 \\ (18.2-83.3)} & \makecell{50.5 \\ (18.2-83.3)} & \makecell{54.8 \\ (18.7-81.8)} & \makecell{52.3 \\ (19.5-78.0)} \\
Sex (Male, Female) & (63.8\%, 36.2\%) & (63.6\%, 36.4\%) & (52.4\%, 47.6\%) & (57.0\%, 43.0\%) \\
\hline
\end{tabular}
\label{table1}
\end{table*}

Table~\ref{table1} presents the distribution of CFP Type 1 data. The gender distribution in the training set shows 63.8\% males and 36.2\% females among migraine sufferers, and 63.6\% males and 36.4\% females among individuals without migraines. In Table~\ref{table2}, which presents CFP Type 2 data, the gender distribution in the training set shows 65.1\% males and 34.9\% females among migraine sufferers, and 65.1\% males and 34.9\% females among non-migraine sufferers. This consistent gender distribution across both data types indicates reliability in demographic representation, which helps mitigate gender bias in the analysis.

Furthermore, Table~\ref{Chi} presenting the Chi-Square test results indicates that there are no significant differences between the training and testing sets across the data types CFP Type 1, CFP Type 2, and OCT ONH. Specifically, for CFP Type 1, the Chi-Square value is 0.148, with 1 degree of freedom and a P-value of 0.929. For CFP Type 2, the Chi-Square value is 0.127, with 1 degree of freedom and a P-value of 0.938. For OCT ONH, the Chi-Square value is 0.063, with 1 degree of freedom and a P-value of 0.969. Since all P-values are much greater than the significance level of 0.05, we fail to reject the null hypothesis, indicating no significant differences between the training and testing sets for these data types. This implies that the distributions of the training and testing sets are similar, ensuring consistent model performance across different datasets.

\begin{table*}[!htb]
\centering
\caption{Distribution of CFP type 2 data [optic nerve head]}
\begin{tabular}{lcccc}
\hline
 & \multicolumn{2}{c}{Train} & \multicolumn{2}{c}{Test} \\
\cline{2-5}
 & Migraine & No Migraine & Migraine & No Migraine \\
\hline
Number of Persons & 69 & 200 & 17 & 50 \\
Number of Images & 444 & 851 & 78 & 195 \\
Age & \makecell{52.8 \\ (17.9-85.0)} & \makecell{51.7 \\ (18.2-86.7)} & \makecell{56.3 \\ (20.7-77.3)} & \makecell{50.9 \\ (19.6-75.8)} \\
Sex (Male, Female) & (65.1\%, 34.9\%) & (65.1\%, 34.9\%) & (51.2\%, 48.8\%) & (59.0\%, 41.0\%) \\
\hline
\end{tabular}
\label{table2}
\end{table*}

\begin{table}[t]
\centering
\caption{Chi-Square Test Results for Training and Testing Sets across Three Data Types}
\begin{tabular}{lccc}
\toprule
 \textbf{Data Type} & $\chi^2$ & \textbf{Degrees of Freedom } & \textbf{P-value} \\
\midrule
CFP Type 1  & 0.148 & 1 & 0.929 \\
CFP Type 2  & 0.127 & 1 & 0.938 \\
OCT ONH     & 0.063 & 1 & 0.969 \\
\hline
\end{tabular}
\label{Chi}
\end{table}



\subsection{Overview of the study methodology}
\noindent Fig.~\ref{fig1} is a schematic representation, illustrating how CFP data and OCT measurement are utilized as  preliminary input data for the model. 

\noindent \textbf{CFP and OCT Data Representation.} Let $I_{\mathrm{CFP}}$ denote the color fundus photography image, represented as a matrix of dimensions $M \times N \times 3$, where $M$ and $N$ correspond to the width and height of the image, and 3 represents the RGB color channels.
Let $M_{\mathrm{OCT}}$ represent the OCT measurements data, defined as a vector of length $K$, where $K$ is the number of OCT measurement parameters.

\noindent \textbf{Feature Extraction by Encoder.} We employed different advanced neural network encoders VGG-16, ResNet-50, Inceptionv3, and Transformer, which are used to convert complex visual inputs into a format suitable for machine learning tasks. These encoders process the CFP images to extract the essential features for accurate classification. Let $f_{\text {encoder }}$ denote the encoder function that transforms the input image $I_{\mathrm{CFP}}$ into a feature vector $F_{\mathrm{CFP}}$:

\begin{equation}
F_{\mathrm{CFP}}=f_{\text {encoder }}\left(I_{\mathrm{CFP}}\right)
\end{equation}
 After the feature extraction process, the OCT measurements data $M_{\mathrm{OCT}}$ (derived from default OCT software) undergoes a specialized linear transformation to distill additional attributes relevant to the condition being examined. The normality of the quantitative data $M_{\mathrm{OCT}}$ was assessed using the Kolmogorov-Smirnov test :
\begin{equation}
F_{\mathrm{OCT}}=T \cdot M_{\mathrm{OCT}}
\end{equation}
where $T$ is a matrix of dimensions $L \times K$, with $L$ representing the length of the feature vector.

\noindent \textbf{Features Concatenation and Classification.}
Then, the feature vectors $F_{\mathrm{CFP}}$ obtained from CFP data and $F_{\mathrm{OCT}}$ from OCT measurements undergo a concatenation process to form a comprehensive feature set, denoted as $F_{\text {combined }}=\operatorname{concat}\left(F_{\mathrm{CFP}}, F_{\mathrm{OCT}}\right)$. This step merges the distinctive yet complementary information derived from both modalities. By combining the textural and visual details captured in the CFP images with the quantitative data from the OCT measurements, the dataset is enriched.

Subsequently, the classifier $f_{\text {classifier }}$ leverages this enriched dataset to predict the likelihood of migraine. 
\begin{equation}
y=f_{\text {classifier }}\left(F_{\text {combined }}\right)
\end{equation}
where $y$ represents the binary outcome, this component of the model utilizes the diverse information encapsulated in $F_{\text {combined }}$. The classifier is designed with a loss function to evaluate prediction accuracy, and through iterative training, refines the model's parameters to minimize prediction error.



\section{Experiment}
In the training configuration for the master model, we utilized an image size of 512$\times$512 pixels. The model was trained over 50 epochs, with each batch consisting of 32 images. For the hardware, we used NVIDIA 3090 servers to facilitate the training process. We also implemented the Adam optimizer to adjust the weights, setting the learning rate at 0.0001. To augment the training data and enhance model robustness, various data augmentation techniques were applied. These included horizontal flipping, panning, scaling, and rotation of the images. This comprehensive training setup was designed to optimize the performance of the deep learning model.

\section{Result}
The evaluation result of the three different models — VGG-16, ResNet-50, and Inceptionv3 — is presented across two CFP type data, both with or without OCT ONH measurements.

We also included a pairwise comparison of all ROC curves, evaluating the differences between the areas under the curves, the associated standard errors, the 95\% confidence intervals for these differences, and the corresponding P-values. If the P-value is less than the conventional threshold of 5\% (P \textless{} 0.05), it indicates a statistically significant difference between the two compared areas.

Continuous variables were expressed as the mean $\pm$ standard deviation (SD), skewed data were expressed as the median (interquartile range, IQR), and categorical variables are reported as percentages.

\begin{table*}[t]
\centering
\caption{Results of Different Model Architectures on CFP data , Based on the Primary Task (Max Sen+Spe). VGG-16 (1), ResNet-50 (1), and Inceptionv3 (1) refer to models trained on CFP Type 1 data , while VGG-16 (2), ResNet-50 (2), and Inceptionv3 (2) refer to models trained on CFP Type 2 data.}
\begin{tabular}{cccccccc}
\toprule
 Model & AUC & Precision & Recall & Specificity & Accuracy & F1  \\
\midrule
 VGG-16 (1)  & 0.858  & 0.782 & 0.669 & 0.889 & 0.807 & 0.721  \\
 & (0.819, 0.897) & (0.735, 0.828) & (0.616, 0.722) & (0.854, 0.924) & (0.763, 0.851) & (0.67, 0.771) \\
VGG-16(2)& 0.675 & 0.556 & 0.449 & 0.824 & 0.700 & 0.496   \\
 & (0.599, 0.75) & (0.477, 0.634) & (0.372, 0.526) & (0.762, 0.885) & (0.627, 0.774) & (0.418, 0.575)  \\  \toprule
ResNet-50 (1) & 0.839 & 0.652 & 0.789 & 0.750 & 0.765 & 0.714  \\
 & (0.798, 0.88) & (0.598, 0.705) & (0.743, 0.835) & (0.701, 0.799) & (0.717, 0.812) & (0.663, 0.765)  \\
 ResNet-50 (2) & 0.697 & 0.569 & 0.474 & 0.824 & 0.709 & 0.517  \\
 & (0.623, 0.771) & (0.491, 0.648) & (0.397, 0.552) & (0.762, 0.885) & (0.636, 0.782) & (0.439, 0.596) \\ \toprule
Inceptionv3 (1) & 0.812 & 0.682 & 0.723 & 0.800 & 0.771 & 0.702  \\
 & (0.768, 0.856) & (0.629, 0.734) & (0.672, 0.773) & (0.755, 0.845) & (0.724, 0.819) & (0.65, 0.753) \\
 Inceptionv3 (2) & 0.731 & 0.495 & 0.692 & 0.654 & 0.667 & 0.578   \\
 & (0.659, 0.802) & (0.417, 0.574) & (0.618, 0.767) & (0.578, 0.73) & (0.591, 0.742) & (0.499, 0.656) \\
\toprule
\end{tabular}
\label{table3}
\end{table*}

\begin{table*}[t]
\centering
\caption{Results of Different Model Architectures on CFP data with OCT ONH, based on the Primary Task (Max Sen+Spe). VGG-16 (1), ResNet-50 (1), and Inceptionv3 (1) refer to models trained on CFP Type 1 data with OCT ONH, while VGG-16 (2), ResNet-50 (2), and Inceptionv3 (2) refer to models trained on CFP Type 2 data with OCT ONH. }
\begin{tabular}{ccccccc}
\toprule
 Model & AUC & Precision & Recall & Specificity & Accuracy & F1 \\
\midrule
 VGG-16 (1) & 0.872 & 0.772 & 0.753 & 0.868 & 0.825 & 0.762  \\ 
 & (0.835, 0.909) & (0.724, 0.819) & (0.704, 0.802) & (0.83, 0.906) & (0.783, 0.868) & (0.714,  0.81) \\
VGG-16 (2) & 0.709 & 0.780 & 0.410 & 0.943 & 0.768 & 0.538  \\
 & (0.636, 0.783) & (0.714, 0.847) & (0.335, 0.485) & (0.907, 0.98) & (0.7, 0.836) & (0.459, 0.617) \\ \toprule
ResNet-50 (1) & 0.853 & 0.726 & 0.717 & 0.839 & 0.794 & 0.721  \\ 
 & (0.813, 0.892) & (0.675, 0.776) & (0.666, 768) & (0.798, 0.88)& (0.748,  0.839)& (0.671, 0.772)\\
ResNet-50 (2) & 0.692 & 0.811 & 0.385 & 0.956 & 0.768 & 0.522 \\
 & (0.618, 0.766) & (0.748, 0.874) & (0.311, 0.458) & (0.923, 0.989) & (0.7, 0.836) & (0.443, 0.6) \\ \toprule
Inceptionv3 (1) & 0.836 & 0.786 & 0.663 & 0.893 & 0.807 & 0.719  \\
 & (0.795, 0.878) & (0.74, 0.832) & (0.609, 0.716)& (0.859, 0.927)& (0.763, 0.851) & (0.668, 0.77) \\
 Inceptionv3 (2) & 0.737 & 0.551 & 0.628 & 0.748 & 0.709 & 0.587  \\
 & (0.666, 0.808) & (0.472, 0.629) & (0.551, 0.706) & (0.678, 0.818) & (0.636, 0.782) & (0.508, 0.665) \\
\toprule
\end{tabular}
\label{table4}
\end{table*}

\subsection{CFP Data Types}

By comparing the model performance on CFP Type 1 and Type 2 data (Table~\ref{table3} and Table~\ref{table4}), it was evident that retinal images with lower magnification, which include the macula, ONH, and vascular arcades, provided more comprehensive insights into retinal changes. Specifically, when using Type 1 data, the VGG-16 model performed with an AUC of 0.872 and an accuracy of 0.825 (Table~\ref{table4}). In contrast, with Type 2 data, the diagnostic performance of all three models significantly decreased, with the highest AUC being 0.737 and the highest accuracy being 0.768 (Table~\ref{table4}). Our initial results suggest that comprehensive retinal images, particularly those that include the macula, ONH, and vascular arcades, could be more effective in revealing pathophysiological changes associated with migraines.

\subsection{Impact of OCT Measurements}

To further analyze the impact of OCT measurements, we compared the performance of the three model with and without OCT ONH measurements according to Table~\ref{table3} and Table~\ref{table4}. For Type 1 data, incorporating OCT measurements improved the VGG-16 model's AUC from 0.858 to 0.872 (+0.014) and accuracy from 0.807 to 0.825 (+0.018), with the F1 score increasing from 0.721 to 0.762 (+0.041). 

A similar trend was observed for Type 2 [ONH] data. Specifically, the AUC rose from 0.675 to 0.709 (+0.034) and accuracy increased from 0.700 to 0.768 (+0.068). These enhancements demonstrate the value of OCT in providing quantitative data that refines the models' capacity to differentiate between individuals with migraines and those without migraines.

\subsection{Statistical Significance} 
\subsubsection{OCT measurements}

\begin{table}[t]
\centering

\caption{Comparison of P-values for Different Models with and without OCT ONH Data  }
\begin{tabular}{lcc}
\toprule
 \textbf{Model}     & \textbf{Without OCT ONH Data} & \textbf{With OCT ONH Data} \\
\midrule
VGG-16 (1)  & Ref   &   0.533    \\
VGG-16 (2)  & Ref   &   0.412    \\
ResNet-50 (1) & Ref  & 0.519  \\
ResNet-50 (2) & Ref  &   0.469\\
Inceptionv3 (1)  &  Ref    &0.541       \\
Inceptionv3 (2)  &  Ref    &  0.359     \\
\hline
\end{tabular}
\label{table5}
\end{table}
Despite the observed improvements with OCT measurements, statistical analysis of the ROC curves (Table~\ref{table5}) indicated that the P-values for comparing models with and without OCT ONH measurements were all above 0.05. This suggests that the differences in performance, while noticeable, are not statistically significant. Therefore, although OCT measurements can enhance model performance, these improvements may not reach statistical significance in this context.

\subsubsection{Model Comparison }

\begin{table}
\centering

\caption{Comparison of P-values for different Backbone Using CFP Data Types 1 and 2, Both Combined with OCT ONH Data}
\begin{tabular}{lcc}
\toprule
 \textbf{Model}     & \textbf{CFP Data Type 1 } & \textbf{CFP Data Type 2 } \\
\midrule
VGG-16   & Ref   &   Ref    \\
ResNet-50 & 0.430   &  0.498 \\
Inceptionv3  &  0.400 &  0.468     \\

\hline
\end{tabular}
\label{table6}

\end{table}

\begin{figure*}
    \centering
    \includegraphics[width=1.04\linewidth]{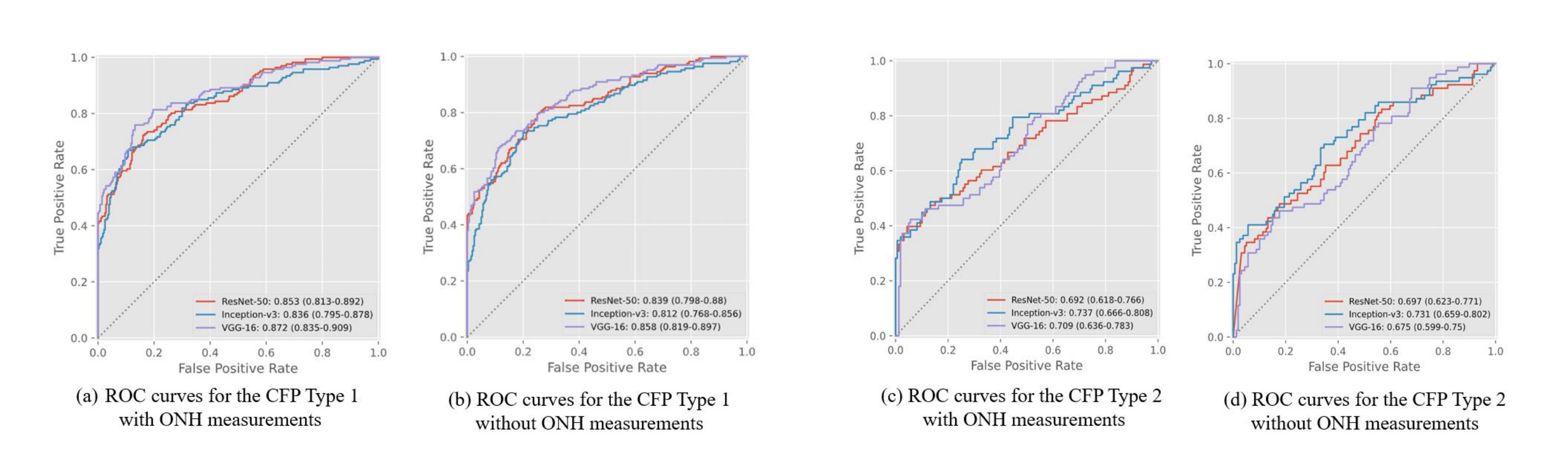}
    \caption{ROC curves for the CFP Type 1 and Type 2 data with and without ONH measurements.}
    \label{fig3}
\end{figure*}

Further comparison of the models using CFP type 1 data revealed that VGG-16's AUC score peaked at 0.872, outperforming ResNet-50 and Inceptionv3, which scored 0.853 and 0.836, respectively (Table~\ref{table4}). VGG-16 also demonstrated the highest accuracy at 0.825, compared to 0.765 for ResNet-50 and 0.771 for Inceptionv3, demonstrating its effectiveness in overall predictions. However, Table~\ref{table6} indicated that these differences were not statistically significant, as P-values were greater than the threshold of 0.05. This suggests that while VGG-16 shows a trend towards better performance, the improvement over ResNet-50 and Inceptionv3 is not significant enough to be deemed statistically meaningful. Figure \ref{fig3} illustrate ROC curves for the CFP Type 1 and Type 2 data with and without ONH measurements.

\subsection{CAMs Visualization}

\begin{figure*}
    \centering
    \includegraphics[width=0.7\linewidth]{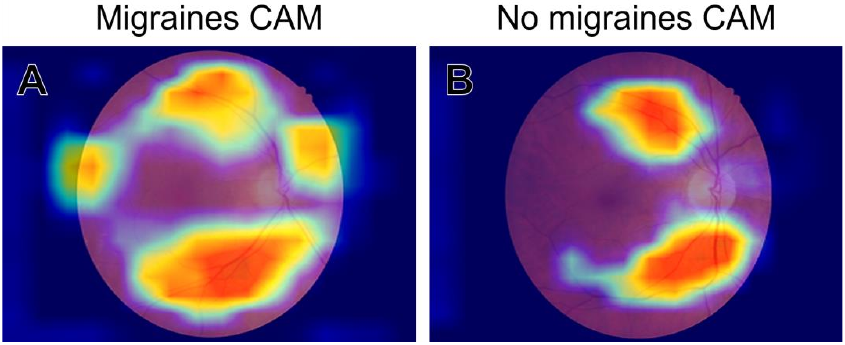}
    \caption{Class Activation Maps (CAM) examples. CAMs generally demonstrated focus at the paravascular arcades. An example is provided for eyes/participants with (A) migraines and (B) no migraines.}
    \label{fig4}
\end{figure*}

The Class Activation Maps (CAMs) provided additional insights into the regions of the retina that the models focused on for making predictions (Fig~\ref{fig4}). CAMs generally exhibited concentrated activity at the paravascular arcades rather than the ONH, particularly in the eyes of participants with migraines (A). This indicates that the paravascular arcades are the most critical regions for identifying retinal changes associated with migraines. 

\section{Discussion}
Currently, two meta-analyses assessing more than 2635 migraine and healthy eyes have explored neuronal alterations associated with migraines~\cite{lin2021retinal, feng2016retinal}. Lin et al. conducted a recent systematic review and meta-analysis, synthesizing findings from twenty-six studies utilizing OCTs to measure the RNFL in both migraine patients and healthy controls~\cite{lin2021retinal}. 
Similarly, Feng et al. analyzed the average RNFL thicknesses from six case-control studies, highlighting a significant thinning of the RNFL in migraine patients across all four quadrants~\cite{feng2016retinal}. These studies provide a robust foundation for understanding the neuronal alterations in migraine sufferers.~\cite{abdelghaffar2022potential}, which would have contributed to the higher AUC achieved using added OCT ONH measurements. However, despite these findings, there is a need for further research to comprehensively validate RNFL thickness as a biomarker for distinguishing migraine from non-migraine patients. Future investigations could explore the use of qualitative OCT data and not just quantitative measures, as the former is not bound by pre-defined software analyses and metrics.

In addition to neuronal alterations, the relationship between migraines and retinal vessel changes has been relatively underexplored \cite{ho2024increased, podraza2024reduction}. Unlu et al. found that retinal artery diameters were increased in patients with $\geq$ five migraine attacks per month compared to the control group~\cite{unlu2017changes}. This dilation of the retinal artery was observed during a migraine attack-free period and was only noticeable in the eye ipsilateral to the headache side~\cite{unlu2017changes}. Furthermore, Asghar et al. demonstrated that the dilation of extra- and intracerebral arteries was associated with migraine attacks~\cite{asghar2011evidence}. These studies shed light on the vascular pathophysiology of migraines and suggest that persistent dilation of the retinal arteries could serve as an observable marker in patients experiencing frequent migraine attacks.

\subsection{Retinal Vascular Pathophysiological Theories}
The exact vascular pathogenesis of migraines remains elusive~\cite{eadie2005pathogenesis}. A theory often put forward is that migraines are caused by the activation of the trigeminovascular system~\cite{mckendrick2022eye}. This leads to the release of pro-inflammatory markers that modulate the diameter of the retinal blood vessels, resulting in vasodilation~\cite{mckendrick2022eye}. Additionally, a review by Intengan et al. stated that low-grade inflammation can cause vascular remodeling in patients with hypertension~\cite{intengan2001vascular}. Pro-inflammatory markers can stimulate the production of angiotensinogen, a molecule that promotes the upregulation of adhesion molecules on endothelial cells lining the arterial walls~\cite{intengan2001vascular, mulvany1999vascular}. This encourages the attachment of leukocytes to the arterial wall endothelium, where they contribute to vascular remodeling~\cite{intengan2001vascular,mulvany1999vascular}. This suggests that temporary changes from the release of pro-inflammatory markers can result in persistent dilation, remodeling, or enlargement of retinal arteries~\cite{mckendrick2022eye,intengan2001vascular}. These alterations in the retinal arteries could serve as a distinguishing biomarker to classify patients with migraines from healthy controls.

\subsection{Significance of Retinal Imaging}
Building on these prior findings, our study aimed to investigate the retinal changes associated with migraines using comprehensive retinal imaging techniques. The comparison between CFP Type 1 and Type 2 data clearly demonstrates that wide-field, low magnification images encompassing the macula, ONH, and vascular arcades provide a richer dataset for model training. The comprehensive images with textual descriptions allowed our models to achieve higher accuracy. This suggests that the inclusion of broader retinal regions captures more relevant features indicative of migraines. The superior performance with CFP Type 1 data indicates that the pathophysiological changes associated with migraines are not localized solely to the ONH but are distributed across the retinal landscape, including the macular and vascular arcade regions.

\subsection{Evaluation of OCT Measurements}
Furthermore, although the inclusion of OCT ONH measurements did not significantly alter the statistical outcomes, the observed improvements in model performance metrics underscore the potential value of OCT data. OCT provides quantitative, high-resolution cross-sectional images of the retina, which complement the qualitative data from CFP and their associated textual descriptions. The slight enhancements in AUC and accuracy with OCT measurements suggest that these quantitative inputs can refine the model's ability to distinguish between normal and migraine-affected retinal structures. However, the lack of statistical significance in these improvements indicates that the added value of OCT measurements may depend on further refinement of the model or a larger sample size.  Besides, qualitative analysis of OCT images may provide valuable and unbounded compared to the quantitative metrics we used.

\subsection{Insights from CAMs}

The consistent identification of the paravascular arcades as critical regions through Class Activation Maps (CAMs) further supports the significance of comprehensive retinal imaging. This insight aligns with the notion that migraines involve widespread vascular and neuronal changes, which are best captured through comprehensive imaging techniques. Therefore, future diagnostic tools and imaging protocols should prioritize these areas when assessing retinal changes in migraine patients.

\section{Future Work}
\subsection{Limitations}
Note that the study is based on a relatively small sample size, as we used a population without retinal disease to avoid confounding factors. This validates the potential of using deep learning models for migraine investigation through retinal analysis. However, the limited data may affect the generalizability of the findings. Additionally, the data used for the diagnosis of migraines were self-reported. There are no objective instrumental or accessible biological markers for diagnosing migraines. Instead, the current gold standard is the diagnosis made by a specialist neurological examination based on the International Classification of Headache Disorders criteria~\cite{wei2023validation}. The criteria for this condition are headache attacks lasting 4-72 hours with specific characteristics, including unilateral location, pulsating quality, moderate to severe pain intensity, aggravation by physical activity, and associated symptoms such as nausea, vomiting, photophobia, and phonophobia~\cite{webpage}.

Secondly, the study population comprised exclusively of eyes without macular or optic nerve head abnormalities, potentially limiting its applicability to real-life scenarios where various concomitant eye diseases may be present. Studies have demonstrated that the diagnostic performance of deep learning models may be enhanced when concomitant eye diseases are present~\cite{cheung2022deep}. This improvement may stem from shared pathophysiological features between the systemic condition and the concomitant eye disease, facilitating the somewhat indirect identification of the systemic condition~\cite{cheung2022deep}. The limitited condition might lead to reduced performance in more complex clinical situations.

\subsection{Direction of Future Research}
While our study demonstrates the potential of using deep learning models with CFP for migraine research, future re-search should explore the integration of additional imaging modalities, such as optical coherence tomography angiography, ultra-widefield imaging, or fluorescein angiography, as well as larger, more diverse datasets to validate and enhance the findings.

Moreover, there is a need to elucidate specific aspects of the retinal vasculature, such as inner and outer diameter, tortuosity, and differences between arterioles and venules, that contribute to distinguishing between participants with and without migraines. Given the black-box nature of machine learning and convolutional neural networks, understanding precisely what features are being processed could significantly advance the interpretability and application of these models.

Additionally, exploring dose-response relationships could provide valuable insights into how different migraine types and severities are reflected in retinal imaging. Understanding these distinctions could refine diagnostic criteria and lead to more tailored monitoring and treatment strategies. 
Future research should also evaluate the performance of the models in external populations with various conditions to optimize the diagnostic capabilities and ensure the models are generalizable across diverse clinical scenarios.

\section{Conclusion}
In conclusion, comprehensive retinal imaging, particularly wide-field CFP data, provides better diagnostic performance for identifying retinal changes associated with migraines compared to CFP imaging localized to ONH. The results demonstrated high discrimination capability using CFP type 1 data, with AUC [95\% CI] ranging from 0.84 [0.80, 0.88] to 0.87 [0.84, 0.91], with no significant differences observed between the VGG-16, ResNet-50, and Inceptionv3 models. However, discrimination using type 2 CFP [ONH] data was reduced, ranging from poor to fair (0.69 [0.62, 0.77] to 0.74 [0.67, 0.81]). Additionally, OCT default ONH measurements did not significantly enhance model performance. Class activation maps (CAMs) indicated that the paravascular arcades were regions of interest. These findings suggest that individuals with migraines exhibit notable microvascular differences rather than significant changes in RNFL thickness. Future research should build on these findings by incorporating larger and more diverse datasets, exploring specific aspects of the retinal vasculature, and evaluating the model's performance in populations with various retinal conditions.


%

\appendices


\section*{Acknowledgment}
We would like to thank the Centre for Eye Health (CFEH), New South Wales, Sydney for providing the data used in this study. We are grateful to the staff at CFEH, particularly the clinicians and technicians who collected and managed the data.

\section*{Conflict of Interest}
The authors declare that the research was conducted in the absence of any commercial or financial relationships that could be construed as a potential conflict of interest.




%

{
    \small
    \bibliographystyle{ieeetr}
    \bibliography{main}

\begin{thebibliography}{10}

\bibitem{abdellatif2018effect}
M.~K. Abdellatif and M.~M. Fouad, ``Effect of duration and severity of migraine on retinal nerve fiber layer, ganglion cell layer, and choroidal thickness,'' {\em European Journal of Ophthalmology}, vol.~28, no.~6, pp.~714--721, 2018.

\bibitem{blair2023rimegepant}
H.~A. Blair, ``Rimegepant: A review in the acute treatment and preventive treatment of migraine,'' {\em CNS drugs}, vol.~37, no.~3, pp.~255--265, 2023.

\bibitem{eigenbrodt2021diagnosis}
A.~K. Eigenbrodt, H.~Ashina, S.~Khan, H.-C. Diener, D.~D. Mitsikostas, A.~J. Sinclair, P.~Pozo-Rosich, P.~Martelletti, A.~Ducros, M.~Lant{\'e}ri-Minet, {\em et~al.}, ``Diagnosis and management of migraine in ten steps,'' {\em Nature Reviews Neurology}, vol.~17, no.~8, pp.~501--514, 2021.

\bibitem{craciun2014long}
L.~Craciun, E.~Gardella, J.~Alving, D.~Terney, I.~Mindruta, J.~Zarubova, and S.~Beniczky, ``How long shall we record electroencephalography?,'' {\em Acta Neurologica Scandinavica}, vol.~129, no.~2, pp.~e9--e11, 2014.

\bibitem{goker2023automatic}
H.~G{\"o}ker, ``Automatic detection of migraine disease from eeg signals using bidirectional long-short term memory deep learning model,'' {\em Signal, Image and Video Processing}, vol.~17, no.~4, pp.~1255--1263, 2023.

\bibitem{ashina2019migraine}
M.~Ashina, J.~M. Hansen, T.~P. Do, A.~Melo-Carrillo, R.~Burstein, and M.~A. Moskowitz, ``Migraine and the trigeminovascular system—40 years and counting,'' {\em The Lancet Neurology}, vol.~18, no.~8, pp.~795--804, 2019.

\bibitem{ray1940experimental}
B.~S. Ray and H.~G. Wolff, ``Experimental studies on headache: pain-sensitive structures of the head and their significance in headache,'' {\em Archives of surgery}, vol.~41, no.~4, pp.~813--856, 1940.

\bibitem{ailani2021atogepant}
J.~Ailani, R.~B. Lipton, P.~J. Goadsby, H.~Guo, R.~Miceli, L.~Severt, M.~Finnegan, and J.~M. Trugman, ``Atogepant for the preventive treatment of migraine,'' {\em New England Journal of Medicine}, vol.~385, no.~8, pp.~695--706, 2021.

\bibitem{haanes2023atogepant}
K.~A. Haanes and L.~Edvinsson, ``Atogepant, the first oral preventive treatment for chronic migraine,'' {\em The Lancet}, vol.~402, no.~10404, pp.~748--749, 2023.

\bibitem{lin2021retinal}
X.~Lin, Z.~Yi, X.~Zhang, Q.~Liu, H.~Zhang, R.~Cai, C.~Chen, H.~Zhang, P.~Zhao, and P.~Pan, ``Retinal nerve fiber layer changes in migraine: a systematic review and meta-analysis,'' {\em Neurological Sciences}, vol.~42, pp.~871--881, 2021.

\bibitem{koruga2024impact}
A.~S. Koruga, T.~Pekmezovi{\'c}, N.~Koruga, A.~Ron{\v{c}}evi{\'c}, S.~Balog, A.~Kokot, D.~Jan{\v{c}}uljak, and S.~B. Soldo, ``The impact of migraine on the thickness of the inner plexiform layer quantifying by optical coherence tomography,'' 2024.

\bibitem{labib2020retinal}
D.~M. Labib, M.~Hegazy, S.~M. Esmat, E.~A.~H. Ali, and F.~Talaat, ``Retinal nerve fiber layer and ganglion cell layer changes using optical coherence tomography in patients with chronic migraine: a case-control study,'' {\em The Egyptian Journal of Neurology, Psychiatry and Neurosurgery}, vol.~56, pp.~1--6, 2020.

\bibitem{feng2016retinal}
Y.-F. Feng, H.~Guo, J.-H. Huang, J.-G. Yu, and F.~Yuan, ``Retinal nerve fiber layer thickness changes in migraine: a meta-analysis of case--control studies,'' {\em Current Eye Research}, vol.~41, no.~6, pp.~814--822, 2016.

\bibitem{reggio2017migraine}
E.~Reggio, C.~G. Chisari, G.~Ferrigno, F.~Patti, G.~Donzuso, G.~Sciacca, T.~Avitabile, S.~Faro, and M.~Zappia, ``Migraine causes retinal and choroidal structural changes: evaluation with ocular coherence tomography,'' {\em Journal of neurology}, vol.~264, pp.~494--502, 2017.

\bibitem{khan2023retinal}
T.~M. Khan, S.~S. Naqvi, A.~Robles-Kelly, and I.~Razzak, ``Retinal vessel segmentation via a multi-resolution contextual network and adversarial learning,'' {\em Neural Networks}, vol.~165, pp.~310--320, 2023.

\bibitem{akinniyi2023multi}
O.~Akinniyi, I.~Razzak, M.~M. Rahman, H.~Sandhu, A.~El-Baz, and F.~Khalifa, ``Multi-classification of retinal diseases using a pyramidal ensemble deep framework,'' in {\em 2023 IEEE International Conference on Image Processing (ICIP)}, pp.~1945--1949, IEEE, 2023.

\bibitem{marziani2009evaluation}
E.~Marziani, P.~Ramolfo, C.~Mariani, S.~Pomati, A.~Giani, M.~Cigada, and G.~Staurenghi, ``Evaluation of retinal nerve fibre layer and ganglion layer cells thickness as biologic marker of alzheimer’s disease,'' {\em Investigative Ophthalmology \& Visual Science}, vol.~50, no.~13, pp.~1096--1096, 2009.

\bibitem{gel2024changes}
M.~S. Gel, A.~Kanat, D.~Seker, H.~Koc, I.~S. Daltaban, H.~Findik, and O.~Lutfi~Gundogdu, ``Changes in retinal nerve fiber layer thickness may be the cause of post-covid-19 headaches,'' {\em Neurological Research}, vol.~46, no.~7, pp.~634--643, 2024.

\bibitem{trinh2024sight}
M.~Trinh, F.~Tang, A.~Ly, A.~Duong, F.~Stapleton, Z.~Ge, and I.~Razzak, ``Sight for sore heads--using cnns to diagnose migraines,'' {\em Investigative Ophthalmology \& Visual Science}, vol.~65, no.~9, pp.~PB0010--PB0010, 2024.

\bibitem{he2016deep}
K.~He, X.~Zhang, S.~Ren, and J.~Sun, ``Deep residual learning for image recognition,'' in {\em Proceedings of the IEEE conference on computer vision and pattern recognition}, pp.~770--778, 2016.

\bibitem{simonyan2014very}
K.~Simonyan and A.~Zisserman, ``Very deep convolutional networks for large-scale image recognition,'' {\em arXiv preprint arXiv:1409.1556}, 2014.

\bibitem{szegedy2016rethinking}
C.~Szegedy, V.~Vanhoucke, S.~Ioffe, J.~Shlens, and Z.~Wojna, ``Rethinking the inception architecture for computer vision,'' in {\em Proceedings of the IEEE conference on computer vision and pattern recognition}, pp.~2818--2826, 2016.

\bibitem{abdelghaffar2022potential}
M.~Abdelghaffar, M.~Hussein, N.~H. Thabet, H.~Elshebawy, L.~I. Daker, and S.~H. Soliman, ``The potential impact of migraine headache on retinal nerve fiber layer thickness,'' {\em The Egyptian Journal of Neurology, Psychiatry and Neurosurgery}, vol.~58, no.~1, p.~141, 2022.

\bibitem{ho2024increased}
K.-Y. Ho, C.-D. Lin, T.-J. Hsu, Y.-H. Huang, F.-J. Tsai, and C.-Y. Liang, ``Increased risks of retinal vascular occlusion in patients with migraine and the protective effects of migraine treatment: a population-based retrospective cohort study,'' {\em Scientific Reports}, vol.~14, no.~1, p.~15429, 2024.

\bibitem{podraza2024reduction}
K.~Podraza, N.~Bangera, A.~Feliz, and A.~Charles, ``Reduction in retinal microvascular perfusion during migraine attacks,'' {\em Headache: The Journal of Head and Face Pain}, vol.~64, no.~1, pp.~16--36, 2024.

\bibitem{unlu2017changes}
M.~Unlu, D.~G. Sevim, M.~Gultekin, R.~Baydemir, C.~Karaca, and A.~Oner, ``Changes in retinal vessel diameters in migraine patients during attack-free period,'' {\em International journal of ophthalmology}, vol.~10, no.~3, p.~439, 2017.

\bibitem{asghar2011evidence}
M.~S. Asghar, A.~E. Hansen, F.~M. Amin, R.~Van Der~Geest, P.~v.~d. Koning, H.~B. Larsson, J.~Olesen, and M.~Ashina, ``Evidence for a vascular factor in migraine,'' {\em Annals of neurology}, vol.~69, no.~4, pp.~635--645, 2011.

\bibitem{eadie2005pathogenesis}
M.~Eadie, ``The pathogenesis of migraine--17th to early 20th century understandings,'' {\em Journal of clinical neuroscience}, vol.~12, no.~4, pp.~383--388, 2005.

\bibitem{mckendrick2022eye}
A.~M. McKendrick and B.~N. Nguyen, ``The eye in migraine: a review of retinal imaging findings in migraine,'' {\em Clinical and Experimental Optometry}, vol.~105, no.~2, pp.~186--193, 2022.

\bibitem{intengan2001vascular}
H.~D. Intengan and E.~L. Schiffrin, ``Vascular remodeling in hypertension: roles of apoptosis, inflammation, and fibrosis,'' {\em Hypertension}, vol.~38, no.~3, pp.~581--587, 2001.

\bibitem{mulvany1999vascular}
M.~J. Mulvany, ``Vascular remodelling of resistance vessels: can we define this?,'' {\em Cardiovascular research}, vol.~41, no.~1, pp.~9--13, 1999.

\bibitem{wei2023validation}
D.~Wei, L.~P. Wong, T.~Loganathan, R.-R. Tang, Y.~Chang, H.-N. Zhou, and M.~K. Kaabar, ``Validation studies on migraine diagnostic tools for use in nonclinical settings: a systematic review,'' {\em Arquivos de Neuro-psiquiatria}, vol.~81, pp.~399--412, 2023.

\bibitem{webpage}
G.~H, ``migraine-without-aura.''
\newblock Accessed: April 18, 2024.

\bibitem{cheung2022deep}
C.~Y. Cheung, A.~R. Ran, S.~Wang, V.~T. Chan, K.~Sham, S.~Hilal, N.~Venketasubramanian, C.-Y. Cheng, C.~Sabanayagam, Y.~C. Tham, {\em et~al.}, ``A deep learning model for detection of alzheimer's disease based on retinal photographs: a retrospective, multicentre case-control study,'' {\em The Lancet Digital Health}, vol.~4, no.~11, pp.~e806--e815, 2022.

\end{thebibliography}
}

\end{document}